\documentclass[a4,12pt]{article}
\newcommand{\nc}{\newcommand}
\nc{\bg}{B. Grzadkowski}
\nc{\non}{\nonumber}
\def\dps{\displaystyle}
\def\mib#1{\mbox{\boldmath $#1$}}

\def\slak{\mbox{$k\!\!\! {/}$}\,}
\def\bra#1{\langle #1 |} \def\ket#1{|#1\rangle}
\def\vev#1{\langle #1\rangle}

\nc{\barx}{\bar{x}}\nc{\pbarn}{\;\hbox {pb}}\nc{\fbarn}{\;\hbox {fb}}
\nc{\hc}{\hbox {h.c.}} \nc{\re}{\hbox {Re}} 
\nc{\mev}{\hbox {MeV}} \nc{\gev}{\;\hbox {GeV}}
\def\gesim{\lower0.5ex\hbox{$\:\buildrel >\over\sim\:$}}
\def\lesim{\lower0.5ex\hbox{$\:\buildrel <\over\sim\:$}}
\nc{\prd}[3]{{\it Phys.\ Rev.}\ {{\bf D{#1}} (#2) #3}}
\nc{\prl}[3]{{\it Phys.\ Rev.\ Lett.}\ {{\bf {#1}} (#2), #3}}
\nc{\plb}[3]{{\it Phys.\ Lett.}\ {{\bf B{#1}} (#2) #3}}
\nc{\npb}[3]{{\it Nucl.\ Phys.}\ {{\bf B{#1}} (#2) #3}}
\nc{\ptp}[3]{{\it Prog.\ Theor.\ Phys.}\ {{\bf {#1}} (#2), #3}}
\nc{\zfp}[3]{{\it Z.\ Phys.}\ {{\bf C{#1}} (#2) #3}}
\nc{\epj}[3]{{\it Eur.\ Phys.\ J.}\ {{\bf C{#1}} (#2) #3}}
\nc{\mpla}[3]{{\it Mod.\ Phys.\ Lett.}\ {{\bf A{#1}} (#2) #3}}
\nc{\rmp}[3]{{\it Rev.\ Mod.\ Phys.}\ {{\bf {#1}} (#2) #3}}
\nc{\ijmpa}[3]{{\it Int.\ J.\ Mod.\ Phys.}\
               {{\bf A{#1}} (#2) #3}}
\nc{\ttbar}{t\bar{t}}         \nc{\bbbar}{b\bar{b}}
\nc{\tanb}{\tan \beta}        \nc{\twbdec}{t\to W^+ b}
\nc{\tbwbdec}{\bar{t}\to W^- \bar{b}}
\nc{\epem}{e^+e^-}            \nc{\eett}{\epem \to \ttbar}
\nc{\sigeett}{\sigma_{e\bar{e}\to\ttbar}}
\nc{\wpwm}{W^+W^-}            \nc{\tbar}{\bar{t}}
\nc{\bbar}{\bar{b}}           \nc{\wpp}{W^+}
\nc{\mt}{m_t}    \nc{\mts}{m_t^2}   \nc{\mw}{m_W}    \nc{\mws}{m_W^2}
\nc{\mz}{m_Z}    \nc{\mzs}{m_Z^2}
\nc{\ttbardec}{\ttbar \to W^+W^-\bbbar}
\nc{\wwbb}{W^+W^-\bbbar}      \nc{\sm}{SM}
\nc{\cw}{\cos\theta_W}        \nc{\sw}{\sin\theta_W}
\nc{\sws}{\sin^2\theta_W}     \nc{\sig}{\sigma_{tot}}
\nc{\lp}{{\ell}^+}              \nc{\lm}{{\ell}^-}
\nc{\epsl}{\epsilon_L}        \nc{\cp}{C\!P}
\nc{\splus}{s_+}       \nc{\smin}{s_-}        \nc{\eps}{\epsilon}
\nc{\psp}{Ps_+}        \nc{\psm}{Ps_-}        \nc{\lsp}{ls_+}
\nc{\lsm}{ls_-}        \nc{\sss}{s_+s_-}      \nc{\m}{m_t}
\nc{\mq}{m_t^2}        \nc{\mr}{\frac{1}{\m}} \nc{\av}{A_{\gamma}}
\nc{\bv}{B_{\gamma}}   \nc{\az}{A_Z}          \nc{\bz}{B_Z}
\nc{\avs}{A_{\gamma}^2}\nc{\azs}{A_Z^2}       \nc{\bzs}{B_Z^2}
\nc{\dav}{\delta \! A_{\gamma}}   \nc{\dbv}{\delta \! B_{\gamma}}
\nc{\dcv}{\delta C_{\gamma}}      \nc{\ddv}{\delta \! D_{\gamma}}
\nc{\daz}{\delta \! A_Z}          \nc{\dbz}{\delta \! B_Z}
\nc{\dcz}{\delta C_Z}             \nc{\ddz}{\delta \! D_Z}
\nc{\dev}{\delta \! E_{\gamma}}   \nc{\dez}{\delta \! E_Z}
\nc{\dfv}{\delta \! F_{\gamma}}   \nc{\dfz}{\delta \! F_Z}
\nc{\rdav}{{\rm Re}(\delta \! A_{\gamma}) \:}
\nc{\rdbv}{{\rm Re}(\delta \! B_{\gamma}) \:}
\nc{\rdcv}{{\rm Re}(\delta C_{\gamma}) \:}
\nc{\rddv}{{\rm Re}(\delta \! D_{\gamma}) \:}
\nc{\rdaz}{{\rm Re}(\delta \! A_Z) \:}
\nc{\rdbz}{{\rm Re}(\delta \! B_Z) \:}
\nc{\rdcz}{{\rm Re}(\delta C_Z) \:}
\nc{\rddz}{{\rm Re}(\delta \! D_Z) \:}
\nc{\idav}{{\rm Im}(\delta \! A_{\gamma}) \:}
\nc{\idbv}{{\rm Im}(\delta \! B_{\gamma}) \:}
\nc{\idcv}{{\rm Im}(\delta C_{\gamma}) \:}
\nc{\iddv}{{\rm Im}(\delta \! D_{\gamma}) \:}
\nc{\idaz}{{\rm Im}(\delta \! A_Z) \:}
\nc{\idbz}{{\rm Im}(\delta \! B_Z) \:}
\nc{\idcz}{{\rm Im}(\delta C_Z) \:}
\nc{\iddz}{{\rm Im}(\delta \! D_Z) \:}
\nc{\cz}{(1+v_e^2)d\:\!'^2}         \nc{\ci}{v_ed\:\!'}
\nc{\ccz}{v_ed\:\!'^2}              \nc{\cci}{d\:\!'}
\nc{\lspace}{\;\;\;\;\;\;\;\;\;\;}  \nc{\llspace}{\lspace \lspace}
\nc{\beq}{\begin{equation}}   \nc{\eeq}{\end{equation}}
\nc{\bea}{\begin{eqnarray}}   \nc{\eea}{\end{eqnarray}}
\nc{\baa}{\begin{array}}      \nc{\eaa}{\end{array}}
\nc{\bit}{\begin{itemize}}    \nc{\eit}{\end{itemize}}
\nc{\ben}{\begin{enumerate}}  \nc{\een}{\end{enumerate}}
\nc{\bce}{\begin{center}}     \nc{\ece}{\end{center}}
\nc{\ocal}{{\cal O}}
\begin{document}
\pagestyle{empty} \setlength{\footskip}{2.0cm}
\setlength{\oddsidemargin}{0.5cm} \setlength{\evensidemargin}{0.5cm}
\renewcommand{\thepage}{-- \arabic{page} --}
\def\mib#1{\mbox{\boldmath $#1$}}
\def\bra#1{\langle #1 |}      \def\ket#1{|#1\rangle}
\def\vev#1{\langle #1\rangle} \def\dps{\displaystyle}
\nc{\tb}{\stackrel{{\scriptscriptstyle (-)}}{t}}
\nc{\bb}{\stackrel{{\scriptscriptstyle (-)}}{b}}
\nc{\fb}{\stackrel{{\scriptscriptstyle (-)}}{f}}
\nc{\pp}{\gamma \gamma}
\nc{\pptt}{\pp \to \ttbar}
   \def\thebibliography#1{\centerline{REFERENCES}
     \list{[\arabic{enumi}]}{\settowidth\labelwidth{[#1]}\leftmargin
     \labelwidth\advance\leftmargin\labelsep\usecounter{enumi}}
     \def\newblock{\hskip .11em plus .33em minus -.07em}\sloppy
     \clubpenalty4000\widowpenalty4000\sfcode`\.=1000\relax}\let
     \endthebibliography=\endlist
   \def\sec#1{\addtocounter{section}{1}\section*{\hspace*{-0.72cm}
     \normalsize\bf\arabic{section}.$\;$#1}\vspace*{-0.3cm}}
\vspace*{-1.7cm}
\noindent

 \vspace{-0.7cm}
\begin{flushright}
$\vcenter{
\hbox{{\footnotesize CERN-PH-TH/2004-050}}
\hbox{{\footnotesize IFT-09-04~~~~FUT-04-01}}
\hbox{{\footnotesize UCRHEP-T372}}
\hbox{{\footnotesize TOKUSHIMA Report}}
\hbox{(hep-ph/0403174)}
}$
\end{flushright}

\vskip 0.6cm
\begin{center}
{\large\bf Optimal-Observable Analysis of Possible New Physics}

\vskip 0.15cm
{\large\bf Using the $\mib{b}$-quark in $\mib{\gamma\gamma\to
t\bar{t}\to bX}$}
\end{center}

\vspace{0.2cm}
\begin{center}
\renewcommand{\thefootnote}{\alph{footnote})}
{\sc Bohdan GRZADKOWSKI$^{\:1),\:}$}\footnote{E-mail address:
\tt bohdan.grzadkowski@fuw.edu.pl},\ \
{\sc Zenr\=o HIOKI$^{\:2),\:}$}\footnote{E-mail address:
\tt hioki@ias.tokushima-u.ac.jp},

\vskip 0.15cm
{\sc Kazumasa OHKUMA$^{\:3),\:}$}\footnote{E-mail address:
\tt ohkuma@ccmails.fukui-ut.ac.jp}\ and\
{\sc Jos\'e WUDKA$^{\:4),\:}$}\footnote{E-mail address:
\tt jose.wudka@ucr.edu}
\end{center}

\vspace*{0.2cm}
\centerline{\sl $1)$ Institute of Theoretical Physics,\ Warsaw
University}
\centerline{\sl Ho\.za 69, PL-00-681 Warsaw, Poland}
\centerline{\sl and}
\centerline{\sl CERN, Department of Physics}
\centerline{\sl Theory Division}
\centerline{\sl 1211 Geneva 23, Switzerland}

\vskip 0.2cm
\centerline{\sl $2)$ Institute of Theoretical Physics,\
University of Tokushima}
\centerline{\sl Tokushima 770-8502, Japan}

\vskip 0.2cm
\centerline{\sl $3)$ Department of Information Science,\
Fukui University of Technology}
\centerline{\sl Fukui 910-8505, Japan}

\vskip 0.2cm
\centerline{\sl $4)$ Department of Physics,\
University of California}
\centerline{\sl Riverside, CA 92521-0413, USA}

\vspace*{0.9cm}
\centerline{ABSTRACT}

\vspace*{0.3cm}
\baselineskip=20pt plus 0.1pt minus 0.1pt
We study possible anomalous top-quark couplings generated by
$SU(2)\times U(1)$ gauge-in\-var\-i\-ant dimension-6 effective
operators, using the final $b$-quark momentum distribution
in $\gamma\gamma\to t\tbar \to bX$. Taking into account
non-standard $t\bar{t}\gamma$, $tbW$ and $\gamma\gamma H$
couplings, we perform an optimal-observable analysis in order
to estimate the precision for the determination of all relevant
non-standard couplings.
\vspace*{0.4cm} \vfill

PACS:  14.65.Fy, 14.65.Ha, 14.70.Bh

Keywords:
anomalous top-quark couplings, $\gamma\gamma$ colliders \\

\newpage
\renewcommand{\thefootnote}{$\sharp$\arabic{footnote}}
\pagestyle{plain} \setcounter{footnote}{0}
\baselineskip=21.0pt plus 0.2pt minus 0.1pt

\sec{Introduction}

Linear colliders of $\epem$ are expected to work as top-quark
factories, and therefore a lot of attention has been paid to
study possible non-standard top-quark interactions through
$e\bar{e}\to t \bar{t}$ (see, for instance,
\cite{Atwood:2001tu,Abe:2001gc} and their reference lists).
An interesting option for such $\epem$ machines could be that
of photon--photon collisions, where initial energetic photons
are produced through electron and laser-light backward
scatterings \cite{Ginzburg:1981vm,Borden:1992qd}.

This type of colliders presents remarkable advantages for the
study of $C\!P$ violation. In the case of $e\bar{e}$ collisions,
the only initial states that are relevant are $C\!P$-even states
$\ket{e_{L/R}\bar{e}_{R/L}}$ under the usual assumption that the
electron mass can be neglected and that the leading contributions
to $\ttbar$ production come from $s$-channel vector-boson
exchanges. Therefore, all $C\!P$-violating observables must be
constructed from final-particle momenta/polarizations. In
contrast, a $\gamma\gamma$ collider offers a unique possibility
of preparing the polarization of the incident-photon beams, which
can be used to construct $C\!P$-violating asymmetries without
relying on final-state information.

This is why a number of authors have considered top-quark
production and decays in $\gamma\gamma$ collisions in order to
study {\it i}) Higgs-boson couplings to the top quark and
photon \cite{Grzadkowski:1992sa}--\cite{Asakawa:2003dh}, or
{\it ii}) anomalous top-quark couplings to the photon
\cite{Choi:1995kp}--\cite{Poulose:1997xk}. However, what is
supposed to be observed in real experiments is combined
signals that originate both from the process of top-quark
production and, {\it in addition}, from its decays. Therefore,
in our latest article \cite{Grzadkowski:2003tf} we
considered $\gamma\gamma \to \ttbar \to {\ell}^+ X$, including
all possible non-standard interactions together (production and
decay), and performed a comprehensive analysis as
model-independently as possible within the effective-Lagrangian
framework of Buchm\"uller and Wyler \cite{Buchmuller:1986jz}.

In this letter, we will carry out an optimal-observable (OO)
analysis, using the final $b$-quark momentum distribution,
as a  complementary work to \cite{Grzadkowski:2003tf}. What we
have to do for this purpose is similar to what has been done
in \cite{Grzadkowski:2003tf}. However, in  the case of the $bX$
final state, we can expect to obtain
independent and valuable information since there is no
branching-ratio suppression for $t\to bW$, in contrast to the
analysis with the final lepton.
One might say that using the final
$b$-quark distribution is not that effective, since the determination of
the $b$-quark momentum is more challenging than that of charged
leptons. However, in any case, it is crucial to tag the final $b$-quark
efficiently in order to distinguish the top-quark production
from the main background of $W^+W^-$ production
\cite{Takahashi:px}. That is, we cannot study top-quark events
without good information on the final $b$-quark, which makes
our analysis realistic.

\sec{Framework}

We use the effective low-energy Lagrangian
\cite{Buchmuller:1986jz,Arzt:gp} to describe possible
new-physics effects. Following this approach, we consider the
Standard-Model (SM) Lagrangian modified by the addition of a
series of $SU(2)\times U(1)$ gauge-invariant operators
${\cal O}_i$ whose coefficients parameterize the low-energy
effects of the underlying high-scale physics.

Since the detailed description of this framework was presented in
\cite{Grzadkowski:2003tf}, we only mention here that the largest
contribution comes from dimension-6 operators, and that these lead to
the following Feynman rules for on-shell
photons, which are necessary for our calculations: \\
(1) $C\!P$-conserving $t\bar{t}\gamma$ vertex
\begin{equation}
\sqrt{2}v \alpha_{\gamma 1}\,\slak\gamma_\mu/{\mit\Lambda}^2,
\end{equation}
(2) $C\!P$-violating $t\bar{t}\gamma$ vertex
\begin{equation}
i\sqrt{2}v \alpha_{\gamma 2}\,
\slak\gamma_\mu \gamma_5/{\mit\Lambda}^2, \label{cpv-vertex}
\end{equation}
(3) $C\!P$-conserving $\gamma\gamma H$ vertex
\begin{eqnarray}
&&-4v \alpha_{h1}\,
\bigl[\:
(k_1 k_2)g_{\mu\nu}-k_{1\nu}k_{2\mu}
\:\bigr]/{\mit\Lambda}^2,
\end{eqnarray}
(4) $C\!P$-violating $\gamma\gamma H$ vertex
\begin{eqnarray}
&&8v \alpha_{h2}\,
k_1^\rho k_2^\sigma \epsilon_{\rho\sigma\mu\nu}/{\mit\Lambda}^2,
\end{eqnarray}
where $v \sim$ 250 GeV, $k$ and $k_{1,2}$ are incoming photon
momenta, and
$\alpha_{\gamma 1,\gamma 2,h1,h2}$ are defined as
\begin{eqnarray}
&&\alpha_{\gamma 1}\equiv
\sin\theta_W{\rm Re}(\alpha_{uW})
+\cos\theta_W{\rm Re}(\alpha'_{uB}),
\\
&&\alpha_{\gamma 2}\equiv
\sin\theta_W{\rm Im}(\alpha_{uW})
+\cos\theta_W{\rm Im}(\alpha'_{uB}),
\\
&&\alpha_{h 1}\equiv
\sin^2\theta_W{\rm Re}(\alpha_{\varphi W})
+\cos^2\theta_W{\rm Re}(\alpha_{\varphi B}) 
-2\sin\theta_W \cos\theta_W{\rm Re}(\alpha_{WB}),
\\
&&\alpha_{h 2}\equiv
\sin^2\theta_W{\rm Re}(\alpha_{\varphi \tilde{W}})
+\cos^2\theta_W{\rm Re}(\alpha_{\varphi \tilde{B}}) 
-\sin\theta_W \cos\theta_W{\rm Re}(\alpha_{\tilde{W}B}),
\end{eqnarray}
$\alpha_i$ and $\alpha_j'$ being the coefficients of
${\cal O}_i$ and ${\cal O}_j'$ ($i=uW$, $\varphi W$, $\varphi B$,
$WB$, $\varphi\tilde{W}$, $\varphi\tilde{B}$, $\tilde{W}B$ and
$j=uB$) respectively, and $\theta_W$ the Weinberg angle. It
will be helpful to note that the SM $f\bar{f}\gamma$
coupling in our scheme is given by $eQ_f \gamma_\mu$,
where $e$ is the proton charge and $Q_f$ is $f$'s electric charge
in $e$ unit (e.g. $Q_u = 2/3$).

The top-quark decay vertex is also affected by some dim-6 operators.
For the on-mass-shell $W$ boson it will be sufficient to consider
just the following $tbW$ amplitude
when $m_b$ is neglected:
\begin{equation}
{\mit\Gamma}^{\mu}_{Wtb}=-\frac{g}{2\sqrt{2}}\:
\bar{u}(p_b)\biggl[\,\gamma^{\mu} (1-\gamma_5)
-{{i\sigma^{\mu\nu}k_{\nu}}\over M_W}
f_2^R (1+\gamma_5) \,\biggr]u(p_t),\label{ffdef}\\
\end{equation}
where $f_2^R$ is given by
\begin{equation}
f^R_2=-\frac{v}{{\mit\Lambda}^2}M_W
  \Bigl[\:\frac{4}{g}\alpha_{uW}
  +\frac12\alpha_{Du}\:\Bigr],  \label{f2R}
\end{equation}
with $\alpha_{Du}$ the coefficient of the operator
${\cal O}_{Du}$~\footnote{Note that there is another potential
    source of contribution to $f^R_2$, which may come from
    ${\cal O}_{\bar D u}$. However, this operator could be eliminated
    using equations of motion; therefore, it is neglected hereafter.
    We thank Ilya Ginzburg for pointing this to us.}.
On the other hand, the $\nu\ell W$ vertex is assumed to receive
negligible contributions from physics beyond the SM.

Finally, the initial-state polarizations are characterized
by the initial electron and positron longitudinal polarizations
$P_e$ and $P_{\bar{e}}$, the average
helicities of the initial-laser photons $P_{\gamma}$ and
$P_{\tilde{\gamma}}$, and their maximum average linear polarizations
$P_t$ and $P_{\tilde{t}}$ with the azimuthal angles $\varphi$ and
$\tilde{\varphi}$
(defined in the same way as in \cite{Ginzburg:1981vm}). The polarizations
$P_{\gamma,t}$ and $P_{\tilde{\gamma},\tilde{t}}$ have to satisfy
\begin{equation}
0 \leq P_{\gamma}^2 + P_t^2 \leq 1,
\ \ \ \ \ \ \ \ \
0 \leq P_{\tilde{\gamma}}^2 + P_{\tilde{t}}^2 \leq 1.
\end{equation}

\sec{Optimal-observable analysis}

The calculation of the cross section is straightforward; to derive
distributions of secondary fermions we have applied
the Kawasaki--Shirafuji--Tsai technique \cite{technique} with
FORM \cite{FORM}  used for the necessary algebraic manipulations.
We neglect contributions that are quadratic in non-standard interactions
and treat the decaying $t$ and $W$ as on-shell particles;
therefore the angular-energy distribution of the $b$ quark
in the $e\bar{e}$ CM frame can be expressed as
\begin{eqnarray}
&&\frac{d\sigma}{dE_b d\cos\theta_b}
=f_{\rm SM}(E_b, \cos\theta_b)
 + \alpha_{\gamma 1} f_{\gamma 1}(E_b, \cos\theta_b)
 + \alpha_{\gamma 2} f_{\gamma 2}(E_b, \cos\theta_b) \non\\
&&\phantom{\frac{d\sigma}{dE_b}}\ \ \
 + \alpha_{h1} f_{h1}(E_b, \cos\theta_b)
 + \alpha_{h2} f_{h2}(E_b, \cos\theta_b)
 + \alpha_d f_d(E_b, \cos\theta_b),
\label{distribution}
\end{eqnarray}
where $f_i(E_b, \cos\theta_b)$ are calculable functions;
$f_{\rm SM}$ denotes the
standard-model contribution, $f_{\gamma 1,\gamma 2}$ describe,
respectively, the anomalous $C\!P$-conserving and
$C\!P$-violating $t\bar{t}\gamma$-vertices contributions,
$f_{h1,h2}$  those generated by the anomalous $C\!P$-conserving and
$C\!P$-violating $\gamma\gamma H$-vertices,
and $f_d$ that by the anomalous $tbW$-vertex  with
\[
\alpha_d = {\rm Re}(f_2^R).
\]
Their analytical form is however too long to be presented in this
letter.

In order to apply the OO technique, we first have to calculate the
following matrix elements
using the weighting functions  $f_i(E_b, \cos\theta_b)$ defined in
eq.~(\ref{distribution}):
\begin{equation}
{\cal M}_{ij}=\int dE_b d\cos\theta_b\,
f_i(E_b, \cos\theta_b)f_j(E_b, \cos\theta_b)/
f_{\rm SM}(E_b, \cos\theta_b),
\end{equation}
and its inverse matrix $X_{ij}$,
where $i,j=1,\cdots, 6$ correspond to SM, $\gamma 1$, $\gamma 2$,
$h1$, $h2$ and $d$ respectively. Then, according to \cite{optimal},
the expected statistical
uncertainty for the measurements of $\alpha_i$ is given by
\beq
|{\mit\Delta}\alpha_i|=\sqrt{I_0 X_{ii}/N_b},
\eeq
where
\[
I_0\equiv \int dE_b d\cos\theta_b\, f_{\rm SM}(E_b, \cos\theta_b)
\]
and $N_b$ is the total number of collected events.

Inverting the matrix ${\cal M}$, we have noticed
that the numerical results for $X_{ij}$
are often unstable: even a tiny variation of ${\cal M}_{ij}$
changes $X_{ij}$ significantly. This indicates that some of
$f_i$ have similar shapes$\,$\footnote{Note that if two $f_i$
    functions were proportional to each other, then the matrix
    ${\cal M}_{ij}$ would have a vanishing determinant, and
    therefore its inverse $X_{ij}$ could not be determined.}\
and therefore their coefficients cannot be disentangled easily.
Indeed, we already encountered a similar trouble in our latest
analysis using final leptons \cite{Grzadkowski:2003tf}.
It is not surprising that we meet this problem again here, since
the main structure of the cross section is determined by that
of $\gamma\gamma \to t\bar{t}$ for both processes.

The presence of such instability forces us to refrain from
determining all the couplings at once through this process alone.
Therefore, hereafter, we assume that some of $\alpha_i$'s have been
measured in other processes (e.g. in
$e\bar{e}\to t\bar{t}\to{\ell}^{\pm}X$).
Fortunately, however, we obtain some
complementary information on coupling constants, which was not
available in our previous analysis \cite{Grzadkowski:2003tf},
where only leptonic distributions were employed.

Below we list all the elements of ${\cal M}\;(={\cal M}^T)$,
which were computed for
\begin{equation}
\sqrt{s_{e\bar{e}}}=500\ {\rm GeV}
\ \ \ {\rm and}\ \ \ {\mit\Lambda}=1\ {\rm TeV}.
\end{equation}

\vskip 0.5cm \noindent
(1) Linear polarization\\
\noindent We chose the following values as typical linear
polarizations: $P_e=P_{\bar{e}}=1$,
$P_t =P_{\tilde{t}}=P_{\gamma}=P_{\tilde{\gamma}}=1/\sqrt{2}$
and $\chi(\equiv \varphi-\tilde{\varphi})=\pi/4$, where
$\varphi$ and $\tilde{\varphi}$ are the azimuthal angles of
$P_t$ and $P_{\tilde{t}}$. They are the same polarizations as
those we used in \cite{Grzadkowski:2003tf}.

\noindent
1-1) $m_H=100$ GeV
\begin{equation}
 \begin{array}{lll}
  {\cal M}_{11}= 0.368\times 10^{2},&  {\cal M}_{12}= 0.787\times 10^{2},&
  {\cal M}_{13}=-0.323\times 10^{1}, \\
  {\cal M}_{14}=-0.145\times 10^{2},&  {\cal M}_{15}=-0.153\times 10^{1},&
  {\cal M}_{16}= 0,\\
  {\cal M}_{22}= 0.169\times 10^{3},&  {\cal M}_{23}=-0.699\times 10^{1},&
  {\cal M}_{24}=-0.299\times 10^{2}, \\
  {\cal M}_{25}=-0.331\times 10^{1},&  {\cal M}_{26}= 0.277\times 10^{1},&
  {\cal M}_{33}= 0.352,             \\
  {\cal M}_{34}= 0.122\times 10^{1},&  {\cal M}_{35}= 0.182,            &
  {\cal M}_{36}=-0.454,             \\
  {\cal M}_{44}= 0.681\times 10^{1},&  {\cal M}_{45}= 0.583,            &
  {\cal M}_{46}= 0.271\times 10^{1}, \\
  {\cal M}_{55}= 0.987\times 10^{-1},& {\cal M}_{56}=-0.281,            &
  {\cal M}_{66}= 0.866\times 10^{1}.
 \end{array}
\end{equation}

\noindent
1-2) $m_H=300$ GeV
\begin{equation}
 \begin{array}{lll}
  {\cal M}_{11}= 0.368\times 10^{2},&  {\cal M}_{12}= 0.787\times 10^{2},&
  {\cal M}_{13}=-0.323\times 10^{1}, \\
  {\cal M}_{14}=-0.359\times 10^{2},&  {\cal M}_{15}=-0.691\times 10^{1},&
  {\cal M}_{16}= 0,\\
  {\cal M}_{22}= 0.169\times 10^{3},&  {\cal M}_{23}=-0.699\times 10^{1},&
  {\cal M}_{24}=-0.742\times 10^{2}, \\
  {\cal M}_{25}=-0.146\times 10^{2},&  {\cal M}_{26}= 0.277\times 10^{1},&
  {\cal M}_{33}= 0.352,             \\
  {\cal M}_{34}= 0.298\times 10^{1},&  {\cal M}_{35}= 0.681,            &
  {\cal M}_{36}=-0.454,             \\
  {\cal M}_{44}= 0.421\times 10^{2},&  {\cal M}_{45}= 0.725\times 10^{1},&
  {\cal M}_{46}= 0.711\times 10^{1}, \\
  {\cal M}_{55}= 0.146\times 10^{1},& {\cal M}_{56}= 0.143,            &
  {\cal M}_{66}= 0.866\times 10^{1}.
 \end{array}
\end{equation}

\noindent
1-3) $m_H=500$ GeV
\begin{equation}
 \begin{array}{lll}
  {\cal M}_{11}= 0.368\times 10^{2},&  {\cal M}_{12}= 0.787\times 10^{2},&
  {\cal M}_{13}=-0.323\times 10^{1}, \\
  {\cal M}_{14}= 0.170\times 10^{2},&  {\cal M}_{15}=-0.101\times 10^{2},&
  {\cal M}_{16}= 0,\\
  {\cal M}_{22}= 0.169\times 10^{3},&  {\cal M}_{23}=-0.699\times 10^{1},&
  {\cal M}_{24}= 0.352\times 10^{2}, \\
  {\cal M}_{25}=-0.206\times 10^{2},&  {\cal M}_{26}= 0.277\times 10^{1},&
  {\cal M}_{33}= 0.352,             \\
  {\cal M}_{34}=-0.148\times 10^{1},&  {\cal M}_{35}= 0.809,            &
  {\cal M}_{36}=-0.454,             \\
  {\cal M}_{44}= 0.935\times 10^{1},&  {\cal M}_{45}=-0.579\times 10^{1},&
  {\cal M}_{46}=-0.283\times 10^{1}, \\
  {\cal M}_{55}= 0.369\times 10^{1},&  {\cal M}_{56}= 0.253\times 10^{1},&
  {\cal M}_{66}= 0.866\times 10^{1}.
 \end{array}
\end{equation}

\noindent
(2) Circular polarization\\
\noindent We took the following values as circular-polarization
parameters: $P_e=P_{\bar{e}}=P_{\gamma}=P_{\tilde{\gamma}}=1$,
which were also used in \cite{Grzadkowski:2003tf}.

\noindent
2-1) $m_H=100$ GeV
\begin{equation}
 \begin{array}{lll}
 {\cal M}_{11}= 0.209\times 10^{2},& {\cal M}_{12}= 0.454\times 10^{2},&
 {\cal M}_{13}= 0, \\
 {\cal M}_{14}=-0.690\times 10^{1},& {\cal M}_{15}=-0.109\times 10^{-3},&
 {\cal M}_{16}= 0, \\
 {\cal M}_{22}= 0.988\times 10^{2},& {\cal M}_{23}= 0        ,&
 {\cal M}_{24}=-0.144\times 10^{2}, \\
 {\cal M}_{25}=-0.227\times 10^{-3},& {\cal M}_{26}= 0.126\times 10^{1},&
 {\cal M}_{33}= 0, \\
 {\cal M}_{34}= 0,                & {\cal M}_{35}= 0        ,&
 {\cal M}_{36}= 0, \\
 {\cal M}_{44}= 0.284\times 10^{1},& {\cal M}_{45}= 0.457\times 10^{-4},&
 {\cal M}_{46}= 0.133\times 10^{1}, \\
 {\cal M}_{55}= 0.739\times 10^{-9},& {\cal M}_{56}= 0.243\times 10^{-4},&
 {\cal M}_{66}= 0.393\times 10^{1}.
 \end{array}
\end{equation}

\noindent
2-2) $m_H=300$ GeV
\begin{equation}
 \begin{array}{lll}
 {\cal M}_{11}= 0.209\times 10^{2},& {\cal M}_{12}= 0.454\times 10^{2},&
 {\cal M}_{13}= 0, \\
 {\cal M}_{14}=-0.178\times 10^{2},& {\cal M}_{15}=-0.177\times 10^{1},&
 {\cal M}_{16}= 0, \\
 {\cal M}_{22}= 0.988\times 10^{2},& {\cal M}_{23}= 0        ,&
 {\cal M}_{24}=-0.373\times 10^{2}, \\
 {\cal M}_{25}=-0.368\times 10^{1},& {\cal M}_{26}= 0.126\times 10^{1},&
 {\cal M}_{33}= 0, \\
 {\cal M}_{34}= 0,                & {\cal M}_{35}= 0        ,&
 {\cal M}_{36}= 0, \\
 {\cal M}_{44}= 0.191\times 10^{2},& {\cal M}_{45}= 0.193\times 10^{1},&
 {\cal M}_{46}= 0.360\times 10^{1}, \\
 {\cal M}_{55}= 0.198,            & {\cal M}_{56}= 0.419,&
 {\cal M}_{66}= 0.393\times 10^{1}.
 \end{array}
\end{equation}

\noindent
2-3) $m_H=500$ GeV
\begin{equation}
 \begin{array}{lll}
 {\cal M}_{11}= 0.209\times 10^{2},& {\cal M}_{12}= 0.454\times 10^{2},&
 {\cal M}_{13}= 0, \\
 {\cal M}_{14}= 0.762\times 10^{1},& {\cal M}_{15}=-0.502\times 10^{1},&
 {\cal M}_{16}= 0, \\
 {\cal M}_{22}= 0.988\times 10^{2},& {\cal M}_{23}= 0        ,&
 {\cal M}_{24}= 0.159\times 10^{2}, \\
 {\cal M}_{25}=-0.105\times 10^{2},& {\cal M}_{26}= 0.126\times 10^{1},&
 {\cal M}_{33}= 0, \\
 {\cal M}_{34}= 0,                & {\cal M}_{35}= 0        ,&
 {\cal M}_{36}= 0, \\
 {\cal M}_{44}= 0.347\times 10^{1},& {\cal M}_{45}=-0.233\times 10^{1},&
 {\cal M}_{46}=-0.138\times 10^{1}, \\
 {\cal M}_{55}= 0.158\times 10^{1},& {\cal M}_{56}= 0.103\times 10^{1},&
 {\cal M}_{66}= 0.393\times 10^{1}.
 \end{array}
\end{equation}
All the elements ${\cal M}_{ij}$ above are given in units of
${\rm fb}$. In these results, the third components of ${\cal M}$
for the circular polarization vanish \cite{Choi:1995kp}. This is
common for analyses using the final lepton and the final $b$-quark.
Also, as in the leptonic case, ${\cal M}_{16}=0$. This time,
however, it is not because of the decoupling, which
holds for the lepton production \cite{Grzadkowski:2002gt}, but
simply because the $tbW$ vertex cannot contribute
to the total cross section of $\gamma\gamma\to t\bar{t}\to bX$,
since
\[
\sigma_{\rm tot}(\gamma\gamma\to t\bar{t}\to bX)
=Br(t\to bX)\sigma_{\rm tot}(\gamma\gamma\to t\bar{t})
\]
and $Br(t\to bX)=1$, whatever anomalous terms are added to the
$tbW$ coupling as long as we assume that a top quark always
decays through $t\to bW$.

When estimating the statistical uncertainty in simultaneous
measurements, e.g. of $\alpha_{\gamma 1}$ and $\alpha_{h 1}$
(assuming all other coefficients are known), we need only
the components with indices 1, 2 and 4. Let us express the
resultant uncertainties as ${\mit\Delta}\alpha_{\gamma 1}^{[3]}$
and ${\mit\Delta}\alpha_{h 1}^{[3]}$, where ``3" shows that we
used the input ${\cal M}_{ij}$, keeping three decimal places.
In order to see how stable the results are, we also computed
${\mit\Delta}\alpha_{\gamma 1}^{[2]}$ and
${\mit\Delta}\alpha_{h 1}^{[2]}$ by rounding ${\cal M}_{ij}$ off
to two decimal places. Then, if both of the deviations
$|{\mit\Delta}\alpha_{\gamma 1,h 1}^{[3]}
-{\mit\Delta}\alpha_{\gamma 1,h 1}^{[2]}|/
{\mit\Delta}\alpha_{\gamma 1,h 1}^{[3]}$ are less than 10\%,
we accept the result as a stable solution.

Although we did not find any stable solution in the
three-parameter analysis, we did find some solutions in a
two-parameter analysis; for those, the numerical results
are presented below. According to the above criterion, the
uncertainties for the following standard deviations
${\mit\Delta}\alpha_{i}$ are limited to 10\%:

\vskip 0.2cm \noindent
1) Linear polarization\\
\noindent$\bullet$ Independent of $m_H$
\begin{equation}
{\mit\Delta} \alpha_{\gamma 2}= 29/\sqrt{N_b},\ \ \ \
{\mit\Delta} \alpha_{d}= 2.6/\sqrt{N_b}, \label{Ri}
\end{equation}
$\bullet$ $m_H=$ 100 GeV
\begin{equation}
{\mit\Delta} \alpha_{h2}= 38/\sqrt{N_b},\ \ \ \
{\mit\Delta} \alpha_{d}= 2.4/\sqrt{N_b},
\end{equation}
$\bullet$  $m_H=$ 300 GeV
\begin{equation}
{\mit\Delta} \alpha_{\gamma 2}= 24/\sqrt{N_b},\ \ \ \
{\mit\Delta} \alpha_{h1}= 2.4/\sqrt{N_b},
\label{g2h1}
\end{equation}
%
\begin{equation}
{\mit\Delta} \alpha_{h1}= 5.4/\sqrt{N_b},\ \ \ \
{\mit\Delta} \alpha_{d}= 4.9/\sqrt{N_b},
\end{equation}
$\bullet$  $m_H=$ 500 GeV
\begin{equation}
{\mit\Delta} \alpha_{\gamma 2}= 23/\sqrt{N_b},\ \ \ \
{\mit\Delta} \alpha_{h1}= 5.0/\sqrt{N_b},
\end{equation}
%
\begin{equation}
{\mit\Delta} \alpha_{h1}= 18/\sqrt{N_b},\ \ \ \
{\mit\Delta} \alpha_{h2}= 22/\sqrt{N_b},
\label{h1h2}
\end{equation}
%
\begin{equation}
{\mit\Delta} \alpha_{h1}= 8.0/\sqrt{N_b},\ \ \ \
{\mit\Delta} \alpha_{d}= 3.3/\sqrt{N_b},
\end{equation}
where $N_b \simeq 18400$ for a luminosity of $L_{e\bar{e}}^{\rm eff}
\equiv \epsilon L_{e\bar{e}}=500\ {\rm fb}^{-1}$ with $\epsilon$
being the relevant detection efficiency and $L_{e\bar{e}}$ being the
integrated luminosity.\footnote{Hereafter we use the tree-level SM formula
    for computing $N_b$, therefore, below we have the same $N_b$ for
    different $m_H$. Also, for illustration, we  assumed
    $L_{e\bar{e}}=500\ {\rm fb}^{-1}$ (adopting $\epsilon=1$) as the
    standard reference point. However, one should not forget that
    tagging a $b$-quark jet including its charge identification is
    harder than that of a lepton.}

\vskip 0.2cm \noindent
2) Circular polarization\\
\noindent$\bullet$ $m_H=$ 100 GeV
\begin{equation}
{\mit\Delta} \alpha_{h1}= 14/\sqrt{N_b},\ \ \ \
{\mit\Delta} \alpha_{d}= 5.2/\sqrt{N_b},
\end{equation}
$\bullet$ $m_H=$ 500 GeV
\begin{equation}
{\mit\Delta} \alpha_{h1}= 10/\sqrt{N_b},\ \ \ \
{\mit\Delta} \alpha_{d}= 4.2/\sqrt{N_b}, \label{Rf}
\end{equation}
where $N_b\simeq 10500$ for $L^{\rm eff}_{e\bar{e}}
=500$ fb$^{-1}$.

It is worth while to compare estimations of sensitivities obtained here
for the $bX$ final
state, with those found in \cite{Grzadkowski:2003tf} in the case of
the ${\ell}^{\pm}X$ final state.
Unfortunately, here, we did not find any stable solution that would allow
for a determination of $\alpha_{\gamma 1}$; the same was also observed
for ${\ell}^{\pm}X$. We therefore have to look
for other suitable processes to determine this parameter.
The precision of $\alpha_{\gamma 2}$ is not very good
either, but it is still much better than in the case of the
lepton analysis. On the other hand, we can see that analyzing
the $b$-quark process with linearly polarized beams enables
us to estimate some ${\mit\Delta}\alpha_i$ that were
unstable in the lepton analysis, i.e.
cases (\ref{g2h1}) and (\ref{h1h2}).
One of them, eq.~(\ref{h1h2}), is especially useful to probe the
$C\!P$ properties of heavy Higgs bosons through the determination of
$\alpha_{h1}$ and $\alpha_{h2}$.
As for the determination of $\alpha_d$, the ${\ell}^{\pm}X$ final state
seems to be more appropriate. These comparisons
show that both final states ($bX$ and ${\ell}^{\pm}X$)
provide complementary information and should therefore be
included in a complete analysis.

The above results are for ${\mit\Lambda}=$ 1 TeV. When one takes
the new-physics scale to be ${\mit\Lambda}'=\lambda{\mit\Lambda}$,
then all the above results (${\mit\Delta}\alpha_i$) are replaced
with ${\mit\Delta}\alpha_i/\lambda^2$, which means that the
right-hand sides of eqs.~(\ref{Ri})--(\ref{Rf}) are multiplied by
$\lambda^2$.

\sec{Summary and Discussion}

We studied here beyond the SM effects in the process $\pp
\to \ttbar \to b X$ for arbitrarily polarized photon beams,
taking advantage of the
fact that polarizations of the incoming-photon beams can be controlled.
Non-standard interactions have been parameterized
through dim-6 local and gauge-symmetric effective operators
\`a la Buchm\"uller and Wyler \cite{Buchmuller:1986jz}. Assuming
that those new-physics effects are small, we have kept only
terms linear in corrections to the SM tree-level vertices.

We applied the optimal-observable technique to final $b$-quark
distributions, and estimated statistical significances of
measuring each (allowed by the gauge invariance) non-standard
parameter. Unfortunately, we had to conclude that it is
never possible to determine all the independent
non-standard parameters at once through $\gamma\gamma\to t\bar{t}\to
b X$ alone. However, we still would be able to perform a
useful analysis if we could utilize the complementary information
collected in other independent processes.

Comments on the background are here in order. The most
serious background is $W$-boson pair productions. Indeed, its
total cross section could be 300 times larger than
$\sigma_{\rm tot}(t\bar{t})$. Fortunately, however, a simulation
study has shown that $t\bar{t}$ events can be selected with a
signal-to-background ratio of 10 by imposing appropriate invariant-mass
constraints on the final-particle momenta \cite{Takahashi:px}.
There, an efficient $b$-quark tagging is crucial, which is a basic
assumption in the analysis presented here.

Some non-standard couplings, which should be determined here,
could also be studied in the standard $e^+e^-$ option of a
linear collider. Therefore, it is worth while to compare the potential
power of the two options. As far as the parameter $\alpha_{\gamma 1}$
is concerned, the $\gamma\gamma$ collider does not allow for its
determination, while it could be determined at $e^+e^-$.
The second $\ttbar\gamma$ coupling $\alpha_{\gamma 2}$,
which is proportional to the real part of the
top-quark electric dipole moment,\footnote{See \cite{Brzezinski:1997av}
    taking into account that the operators
    $\ocal_{uB}$, $\ocal_{qB}$ and $\ocal_{qW}$ are redundant.}
can be measured here. It should be recalled that energy and
polar-angle distributions of leptons and $b$-quarks in $e^+e^-$
colliders are sensitive only to the imaginary part of the electric
dipole moment,\footnote{However, it should be emphasized that there
    exist observables sensitive also to the real part of the top-quark
    electric dipole moment, see \cite{Chang:1993fu}.}
while here the real part could be determined.
For the measurement of $\gamma\gamma H$
couplings, $e^+e^-$ colliders are, of course, useless, while here,
for the $bX$ final state both $\alpha_{h 1}$ and $\alpha_{h 2}$
could be measured. In the case of the decay form factor $\alpha_d$
measurement, the $e^+e^-$ option seems to be a little more advantageous,
especially if $e^+e^-$ polarization can be tuned
appropriately~\cite{Grzadkowski:1996kn}.

It should be emphasized here that the effective-operator
strategy adopted in this article is valid only for ${\mit\Lambda}
\gg v\simeq 250 \gev$, in contrast to the analysis of
$\epem \to t\bar{t}\to {\ell}^{\pm}X$ performed in
\cite{Grzadkowski:2002gt} and \cite{Grzadkowski:1996kn}  for example.
Should the reaction $\gamma \gamma \to \ttbar \to b X$
exhibit a deviation from the SM predictions  that cannot be
described properly within this framework, we would have an
indication of low-energy beyond-the-SM physics, e.g.
two-Higgs-doublet models with new
scalar degrees of freedom of relatively low mass scale.

\vspace{0.6cm}
\centerline{ACKNOWLEDGMENTS}

\vspace{0.3cm}
One of us (Z.H.) would like to thank Tohru Takahashi and
Katsumasa Ikematsu for very useful correspondence. This
work is supported in part by the State Committee for
Scientific Research (Poland) under grant 1~P03B~078~26 in
the period 2004--2006 and the Grant-in-Aid for Scientific
Research No.13135219 from the Japan Society for the
Promotion of Science.

\vspace*{0.8cm}

\end{document}